\documentclass[twocolumn, tighten, times]{aastex631}

\usepackage{xcolor}
\usepackage{colortbl}
\usepackage{color}
\graphicspath{{./}{figures/}}
\usepackage{hyperref}
\usepackage{float}
\usepackage{graphicx}
\begin{document}

\title{Unraveling the Nature of HAWC J1844-034 with Fermi-LAT Data Analysis and Multi-wavelength Modeling}

\author[0009-0004-1305-9578]{Sovan Boxi}

\email{sovanboxi@rrimail.rri.res.in}
\affiliation{Raman Research Institute \\
C. V. Raman Avenue, 5th Cross Road, Sadashivanagar, Bengaluru, Karnataka 560080, India}
\author[0000-0001-8585-9743]{Saptarshi Ghosh}
\email{ghosh@apc.in2p3.fr}
\affiliation{Astroparticule et Cosmologie, Université Paris Cité, Paris 75013, France}
\author[0000-0002-1188-7503]{Nayantara Gupta}
\affiliation{Raman Research Institute \\
C. V. Raman Avenue, 5th Cross Road, Sadashivanagar, Bengaluru, Karnataka 560080, India}
\email{nayan@rri.res.in}

\begin{abstract}
The extended ultra-high-energy (UHE) gamma-ray source HAWC J1844-034 is closely associated with two other sources, HAWC J1843-032 and HWC J1846-025. Moreover, other gamma-ray observatories like H.E.S.S., LHAASO, and Tibet AS$_{\gamma}$ have detected UHE gamma-ray sources whose spatial positions coincide with the position of HAWC J1844-034. The UHE gamma-ray data from several observatories help analyse the spectral features of this source in detail at TeV energies. Of the four pulsars near HAWC J1844-034, PSR J1844-0346 is closest to it and possibly supplies the cosmic-ray leptons to power this source. We have analysed the Fermi-LAT data to explore this source's morphology and identify its spectral feature in the Fermi-LAT energy band. After removing the contribution of the pulsar to the gamma-ray spectral energy distribution by pulsar phased analysis, we have obtained upper limits on the photon flux and identified the GeV counterpart PS J1844.2-0342 in the Fermi-LAT energy band with more than 5$\sigma$ significance, which may be a pulsar wind nebula (PWN). Finally, the multi-wavelength spectral energy distribution is modeled, assuming HAWC J1844-034 is a PWN.

\end{abstract}


\section{Introduction} \label{sec:intro}
The ultra-high-energy (UHE) gamma-ray window has opened up a great opportunity to study the highest-energy cosmic accelerators in the Galaxy. The High Altitude Water Charenkov gamma-ray observatory (HAWC) detects gamma rays and cosmic rays of energies ranging from 100 GeV to beyond 100 TeV. The third HAWC catalogue \citep{Albert_2020} of very high energy (VHE) gamma-ray sources has a list of 65 sources detected at more than 5$\sigma$ significance. The Large High Altitude Air Shower Observatory (LHAASO) released its first source catalogue in 2024 which reports about 43 UHE gamma-ray sources \citep{Cao_2024}. Observations of TeV gamma-ray sources by the High Energy Spectroscopic System (H.E.S.S.) revealed that most TeV gamma-ray sources are spatially extended sources. Due to their association with supernova remnants, molecular clouds and pulsars, it is nontrivial to identify the origin of the extended emissions.

\begin{figure*}[ht]
\centering
\includegraphics[width=0.65\textwidth]{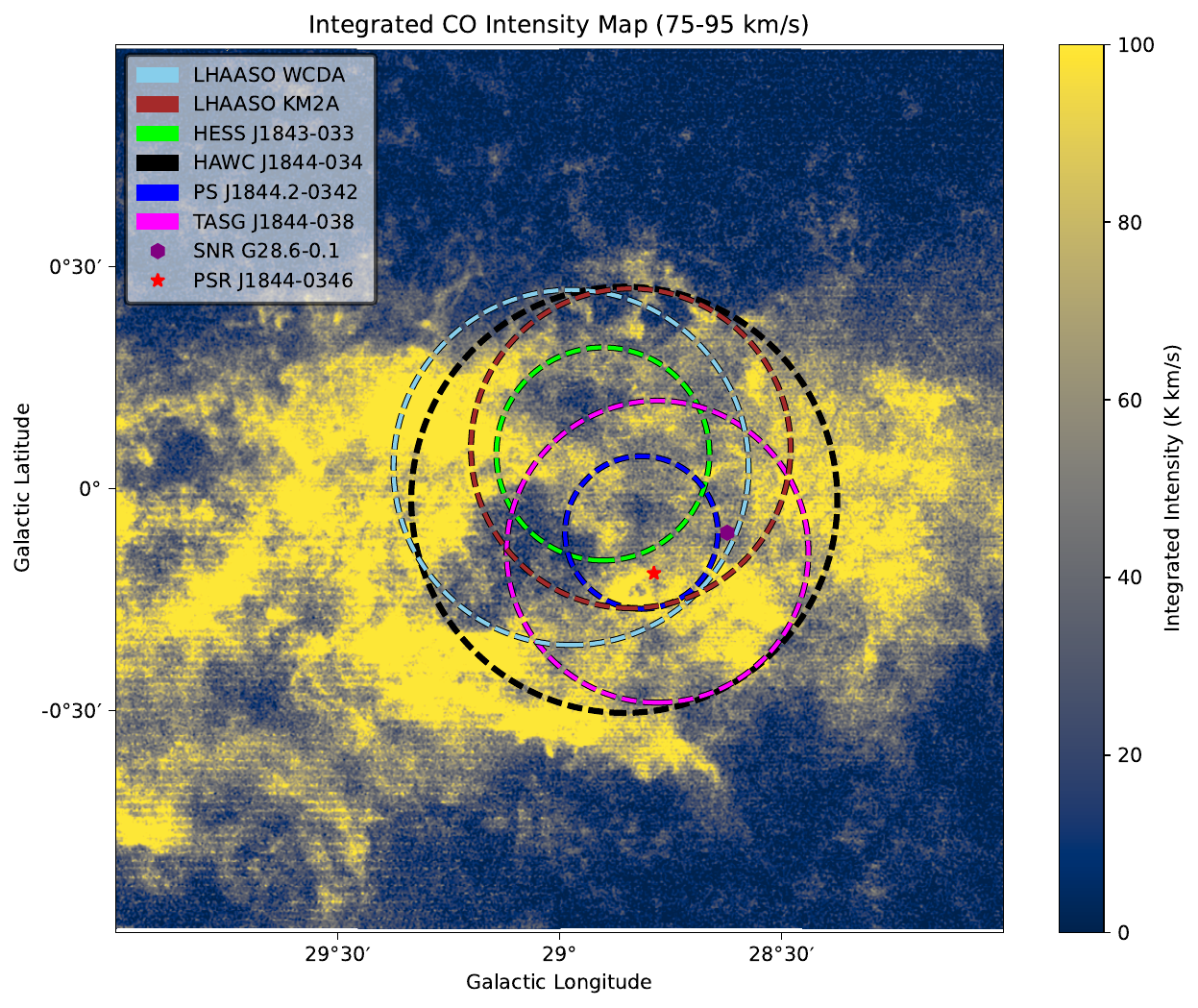}
\caption{$2^{\circ}\times 2^{\circ}$ FUGIN survey molecular cloud distribution in the HAWC J$1844$-$034$ region of the integrated velocity channel from $75${-}$95$ km/s investigated with $^{12}$CO($J=1{-}0$) line emission using FUGIN public data \citep{10.1093/pasj/psx061}. 
    $1\sigma$ extension contours from H.E.S.S.\citep{HESScollab2018ana}, HAWC \citep{Albert_2023}, Fermi (this work), and TS-$\gamma$ \citep{Tibet} experiment counterparts are indicated in the left legend of the figure. LHAASO 39\% extension radius measurement from LHAASO KM2A and WCDA is represented in brown and sky blue, respectively \citep{Cao_2024}. Positions of PSR J1844-0346 \citep{Manchester_2005} and SNR G$28.6$-$0.1$ \citep{2001PASJ...53L..21B,2003ApJ...588..338U} are marked in red and purple, respectively. 
}
\label{fig:morphology}
\end{figure*}

\par 
The H.E.S.S. Galactic Plane Survey (HGPS) detected HESS J1843-033, a gamma-ray source with no clear association, emitting up to 30 TeV photons \citep{HESScollab2018ana}. Subsequent observations by HAWC revealed two sources, 2HWC J1844-032 and eHWC J1842-035, near the H.E.S.S. detection. These sources were associated with emissions extending up to 56 TeV (\cite{Abeysekara:2021DF}). The LHAASO experiment identified LHAASO J1843-0338, which emitted photons exceeding 100 TeV \citep{UHELHAASO}, which is a potential counterpart of HESS J1843-033 and 2HWC J1844-032. \cite{Tibet}, using the Tibet AS-$\gamma$ experiment, reported TASG J1844-038, a gamma-ray source located near HESS J1843-033. They detected emissions above 25 TeV with a significance of 6.2$\sigma$ and discussed two possible associations: the pulsar PSR J1844-0346 and the supernova remnant (SNR) G28.6-0.1. Earlier, \cite{1989ApJ...341..151H} identified a radio complex near the region, including areas of non-thermal emission. Subsequent X-ray observations with Chandra and ASCA confirmed non-thermal radiation from high-energy electrons, consistent with a shell-type SNR \citep{2001PASJ...53L..21B,2003ApJ...588..338U}. This suggested a possible hadronic origin for the gamma rays, supported by the source's extended diffuse structure and compatibility with the SNR's age. The alternative leptonic scenario involving a pulsar wind nebula (PWN) powered by PSR J1844-0346 has also been proposed. \cite{Sudoh_2021} emphasised that the flat spectral index of TASG J1844-038 and the energetics of PSR J1844-0346 align with a TeV PWN interpretation. They also qualitatively discussed an inverse Compton scattering scenario and its potential correlation with the extended VHE source.



\par 
The unidentified source HESS J1843-033 is modelled as a large extended Gaussian component resulting from merging two Gaussian components previously detected by the HGPS pipeline. The radio SNR G28.6-0.1 and the energetic pulsar PSR J1844-0346 are located within one of the previously detected Gaussian components.
Figure 7 of \cite{PSRJ1844Distance} shows the region's complexity and a peak significance at nearly $6\sigma$ near the position of the synchrotron emitting SNR G28.6-0.1.
PSR J1844-0346 lies approximately $10'$ from the centre of SNR G28.6-0.1 \citep{10.1093/mnras/sty359}. \cite{PSRJ1844Distance} investigated the region and found no evidence of radio or X-ray emissions indicative of a bright synchrotron PWN. They also argued that an association between this pulsar and SNR G28.6-0.1 is unlikely, even with the most optimistic distance estimates for G28.6-0.1. Given the characteristic age of PSR J$1844$-$0346$ and the decay of its magnetic field \citep{2010ApJ...715.1248T}, synchrotron emission is expected to be faint and challenging to detect. However, MAGPIS radio and Spitzer infrared data revealed extended bright emission southeast of the pulsar, potentially linked to the star-forming region N49 \citep{2012AJ....144..173D}, embedded within the TASG extension \citep{PSRJ1844Distance,Tibet}.

\par Very recently, LHAASO detected the UHE gamma-ray source 1LHAASO J1843-0335u at an angular separation of 0.06$^{\circ}$ \citep{Cao_2024} from HESS J1843-033. It has been detected above 100 TeV with a TS value of 295.8 by the KM2A component of LHAASO.

\par A detailed observational data analysis of the source region of eHWC identified three components: HAWC J1844-034, HAWC J1843-032 and HAWC J1846-025 \citep{Albert_2023}. The most significant and extended source among these three is HAWC J1844-034. We have analysed the Fermi-LAT data from the source region to search for the emission from the extended source HAWC J1844-034. After pulsar phased analysis of the Fermi-LAT data, we eliminated the gamma-ray emission from the pulsar and obtained upper limits from HAWC J1844-034. Finally, after combining the observational data from various observations, we have modelled the spectral energy distribution, assuming the extended emission originates from the PWN powered by PSR J$1844$-$0346$.

\section{HAWC J1844-034}
HAWC J1844-034 is a bright extended gamma array source in the eHWC J1842-035 region emitting photons at energies up to 175 TeV \citep{Albert_2023}. The source exhibits an extended radially symmetric Gaussian morphology and has been detected with a $\sim 26\sigma$ significance. Its best fit position is RA(J2000) = $281.02^{+0.05}_{-0.05}$, Dec(J2000) = $-3.64^{+0.05}_{-0.004}$ with  1$\sigma$ extension radius of $0.48^{+0.02}_{-0.02}$ \citep{Albert_2023}.
\par 
\par Several VHE sources have been identified by different observatories in the HAWC region, as illustrated in Fig.~\ref{fig:morphology}. For instance, LHAASO detection of ultra-high energy source LHAASO J1843-0338 has an angular separation of $0.27^{\circ}$ from the HAWC centroid with an extension of $0.3^{\circ}$. Tibet As-$\gamma$ experiments have observed another VHE source TASG J1844-038 localised in that region with an Offset of $0.14^{\circ}$ from the HAWC centroid and have an extension of $0.34^{\circ}$. H.E.S.S. HGPS survey found another interesting source, HESS J1843-033, near the HAWC centre at an angular separation of $0.11^{\circ}$. Such spatial overlap of detections across multiple gamma-ray observatories highlights their excellent angular resolution at high energies \citep{Albert_2023}. Despite these detections, the exact association for HAWC J1844-034 remains uncertain. Further investigation on this emission region's kinematical and morphological properties is required.

\par A possible association with the supernova remnant G$28.6$-$0.1$ has been proposed. The shell-type SNR observed by \cite{2001PASJ...53L..21B, 2003ApJ...588..338U} lies $0.28^{\circ}$ away from HAWC J1844-034 centeroid. It appears to interact with the molecular clouds in a velocity channel of 86 km/s with an estimated distance of $9.6 \pm 0.3$ kpc \citep{10.1093/mnras/sty817}. The molecular clouds distribution from $75$-$95$ km/s channel investigated with $^{12}$CO($J=1{-}0$) emission line indicates the overlap of the TASG source with the SNR (Fig.~\ref{fig:morphology}), suggesting that SNR G$28.6-0.1$ could be a possible PeVatron candidate \citep{Tibet}. However, modeling efforts from the HAWC collaboration suggest the estimated hadron spectrum has a smaller cut-off energy than the expected PeVatron cut-off energy \citep{Albert_2023}. The gradient in the molecular cloud from our Fig.~\ref{fig:morphology} doesn't reflect the observed gamma-ray emission in the HAWC significance map (see Fig.~1 of \cite{Albert_2023}), making the hadronic scenario more challenging. 
 
 \par
 
 There are at least three pulsars within the best-fit size of HAWC J1844-034: PSR J1843-0355 ($\dot{E} = 1.77 \times 10^{34}$ erg s$^{-1}$, $d \sim 5.8$ kpc), PSR J1844-0310 ($\dot{E} = 2.79 \times 10^{33}$ erg s$^{-1}$, $d \sim 6$ kpc), and PSR J1844-0346 ($\dot{E} = 4.2 \times 10^{36}$ erg s$^{-1}$, $d \sim 4.3 /2.4$kpc) \citep{Manchester_2005,PSRJ1844Distance, Wu_2018}. Also, PSR J1841-0345 ($\dot{E} = 2.7 \times 10^{35}$ erg s$^{-1}$, $d \sim 3.8$ kpc) is suggested as a possible counterpart source to LHAASO J1843-0338 in \cite{Wilhelmi_2022} but it is located $0.63^{\circ}$ away from HAWC J$1844$-$034$, placing it outside of the extension of HAWC J$1844$-$034$. Therefore, we rule out PSR J$1841$-$0345$ from the counterpart source candidates. Among the remaining pulsars, PSR J$1843$-$0355$
and PSR J1844-0310 are both disfavored due to their low spin-down power, which is insufficient to account for the gamma-ray luminosity observed in the HAWC J1844-034. Finally, PSR J1844-0346 has a spin-down power approximately two orders of magnitude higher than the gamma-ray luminosity. This GeV pulsar was discovered in the Fermi-LAT blind search \citep{Clark_2017}. The Fermi-LAT catalogue identifies this as 4FGL J$1844.4$-$0345$ \citep{Abdollahi_2020}. The exact distance of the pulsar is still unknown. The empirical estimates assuming gamma-ray luminosity scale with spin-down luminosity as $\sqrt{\dot{E}}$  \citep{Wu_2018} suggest the distance to be 2.4 kpc \citep{zhang2024pwn}. Alternatively, gamma-ray pulsar distance estimate based on the empirical formula \citep{2010ApJ...725..571S} suggests the pseudo distance to be 4.3 kpc. Moreover, it is located closest to the best-fit positions of HAWC J1844-034, HESS J1843-033, and TASG J1844-038 with the angular distances of $0.18^\circ$, $0.28^\circ$, and $0.05^\circ$, respectively.

 \par Although leptonic and hadronic emission mechanisms associated with the SNR  have been explored in the HAWC collaboration \citep{Albert_2023} modeling effort, several questions remain unanswered. X-ray data were not incorporated to constrain the magnetic field, and Fermi-LAT data were not included in spectral fits. The authors qualitatively discussed the possibility of a PWN powered by PSR J1844-0346 but no direct modeling of the pulsar scenario was performed.

\section{Fermi Data Analysis}
\subsection{Standard Analysis}
We performed a detailed analysis of Fermi-LAT using 16 years of data, from MET $239557417$ ($2008$-$08$-$04$ $15$:$43$:$36$ UTC) to MET $739053099$ ($2024$-$06$-$02$ $20$:$31$:$34$ UTC) in a region of interest (ROI) of $12^{\circ} \times 12^{\circ}$ centred on the HAWC J1844-034. We referred to HAWC J1844-034 as the target source while pursuing our standard analysis. The Fermi-LAT point spread function (PSF) improved significantly above 1 GeV, achieving an angular resolution of $0.1^{\circ}$ above 10 GeV. So, we restricted our analysis to the  1-500 GeV to leverage the better PSF at higher energies and to reduce the contamination from the flagged low-energy sources and molecular gas clumps.

The analysis used the latest comprehensive $4$FGL-DR$4$ source catalogue along with the Galactic diffused template \texttt{gll\_iem\_v07.fits} and isotropic template \texttt{iso\_P8R3\_SOURCE\_V3\_v1.txt}.
We have adopted the \texttt{P8R3\_SOURCE\_V3} instrument response function with spatial binning of $0.05^\circ$ and eight energy bins per decade. Energy dispersion corrections were applied in all analyses except for the diffuse isotropic background. Photon events were categorised by their angular reconstruction quality into 4 PSF classes (PSF0, PSF1, PSF2, PSF3), where PSF0 represents the worst quality and PSF3 the best. Likelihood analysis was performed for each PSF class separately, and the results were combined into a global likelihood function representing all the events in the ROI. We excluded the photons detected at the zenith angles $<90^{\circ}$ to minimise the contamination from the Earth limb gamma rays caused by cosmic ray interactions.

The \texttt{gta.optimize()} tool was executed two or three times, allowing free the normalisation of the parameters of sources in ROI to vary to ensure convergence near the global likelihood maxima. The \texttt{gta.print\_roi()} output identified weakly significant sources. We removed non-detected sources with $TS \leq 3$ or $N_{\text{pred}} \leq 3$ to refine the model and improve the fit. The presence of numerous sources near the target source required a careful fitting approach. Nearby sources within $3^\circ$ were handled by freeing their normalisation parameters within $3^\circ$ of the target while keeping other parameters fixed to mitigate PSF-induced overlaps. High-TS sources and diffuse galactic and isotropic background normalisations were also freed to ensure accurate modeling of the target source. Finally, we performed a fit using  \texttt{gta.fit()} to fit all the parameters within the ROI iterating until the fit quality reaches $3$. We used the \texttt{NEWMINUIT} optimiser for this purpose. The \texttt{NEWMINUIT} optimiser returned the best-fit model and flagged cases where convergence was problematic. The fit will give us the best possible model based on the input data. Diagnostic plots were inspected to confirm the fit quality. Then we employed \texttt{gta.localize()} and \texttt{gta.find\_sources()} algorithms to refine the unmodeled regions. We have discussed the relevant outputs in the Results section.

\begin{figure}[!htbp]
    \centering
    \includegraphics[width=0.35\textwidth]{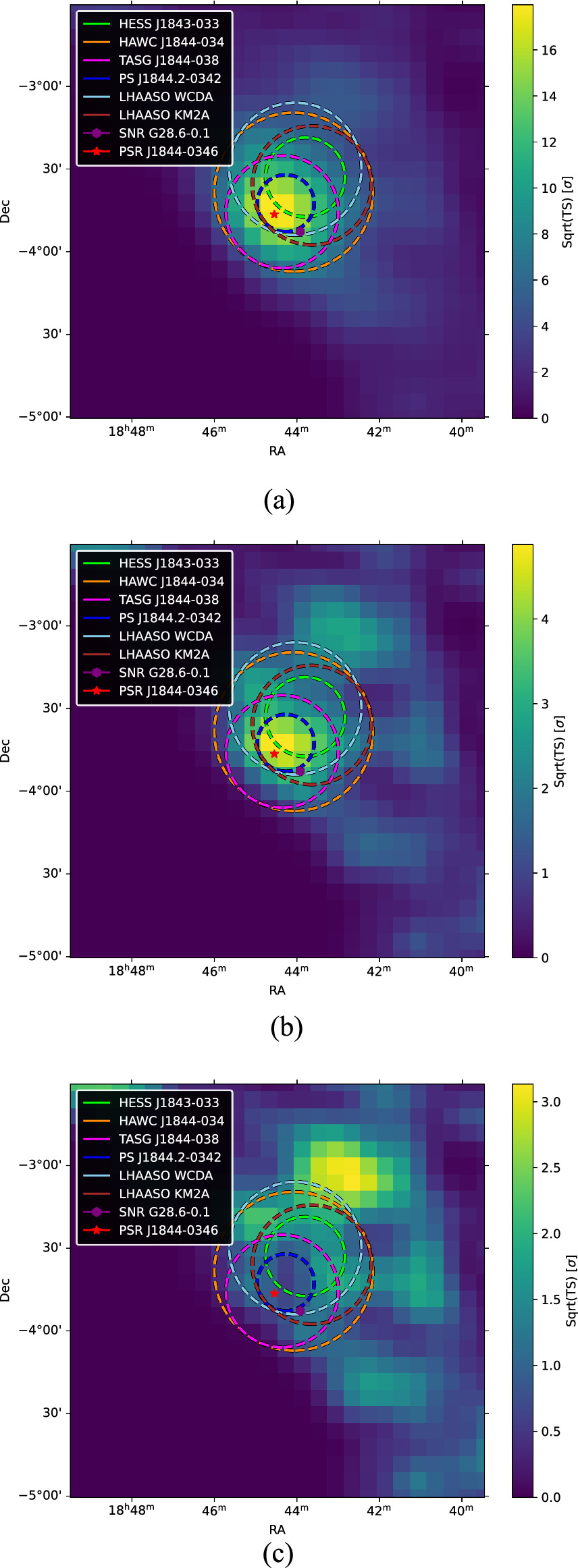}
    \caption{$\sqrt{\mathrm{TS}}$ maps from pulsar phased analysis in the full energy range considered around the HAWC J1844-034 source.
       (a) In the top panel, the On-phase $\sqrt{\mathrm{TS}}$ map is shown.
       (b) In the middle panel, the Off-phase $\sqrt{\mathrm{TS}}$ map is shown before looking for new sources through \texttt{gta.find\_sources()} tool.
       (c) In the bottom panel, we show $5\sigma$ level reduction in the residuals at the position of the pulsar after modeling the new source PS J1844.2-0342. $1\sigma$ extension contours from H.E.S.S.\citep{HESScollab2018ana}, HAWC \citep{Albert_2023}, PS J1844.2-0342 (this work), and TS-$\gamma$ \citep{Tibet} experiment counterparts are indicated in the left legend of the figures. LHAASO 39\% extension radius measurement from LHAASO KM2A and WCDA is represented in brown and sky blue, respectively \citep{Cao_2024}. Positions of PSR J1844-0346 \citep{Manchester_2005} and SNR G$28.6$-$0.1$ \citep{2001PASJ...53L..21B,2003ApJ...588..338U} are marked in red and purple, respectively.}
    \label{fig:tsmaps}
\end{figure}

\subsection{Pulsar-phased Analysis}
We analysed 11 years of Fermi-LAT data collected between August 2008 and December $2019$, focusing on a $10^{\circ} \times 10^{\circ}$ ROI around the pulsar $4$FGL J$1844.4$-$0345$. The analysis used Pass 8 data in the \texttt{Source} class (\texttt{evclass=128}) and \texttt{FRonT+BACK} type (\texttt{evtype=3}) events, covering the energy range from $100$ MeV to $600$ GeV. Event data files containing the \texttt{PULSE\_PHASE} column were obtained directly from the Fermipy Pulsar catalogue\footnote{FermiLAT website at \url{https://fermi.gsfc.nasa.gov/ssc/data/access/lat/3rd_PSR_catalog/3PC_HTML/J1844-0346.html}} while the spacecraft file was downloaded from the Fermi-LAT data server. After having these files, we followed the analysis procedure given in the pulsar-phased Analysis tutorial\footnote{Fermipy documentation at \url{https://fermipy.readthedocs.io/en/latest/notebooks/phase_analysis.html}}. To limit contamination from Earth's limb gamma-ray emission, events with a zenith angle greater than $105^{\circ}$ were excluded. We modelled the region using all point-like and extended sources listed in the $4$FGL-DR$4$ catalogue within a $12^{\circ} \times 12^{\circ}$ area, including galactic diffuse (\texttt{gll\_iem\_v07.fits}) and isotropic (\texttt{iso\_P8R3\_SOURCE\_V3\_v1.txt}) background templates. Following the Fermipy pulsar-phased analysis framework, we conducted a joint analysis of On-phase and Off-phase data.
First, we employed \texttt{gta.optimize()} to refine the fit parameters by ensuring convergence to the global likelihood function. Subsequently, we performed the initial fit, allowing the Galactic diffused emission, isotropic background and sources with $TS>10$ to vary. Diagnostic plots revealed positive residuals near the pulsar's position and negative residuals within the three degrees of ROI. We freed the normalisation parameters of the sources of the affected region, including the target source and background normalisation to improve the fit. A refit improved the ROI significantly, leaving no regions with $TS\sim 5\sigma$ except at the position of Pulsar (see Fig.~\ref{fig:tsmaps}(b)). The TS value, which quantifies the source detection significance, is defined as:
\begin{equation}
 \mathrm{TS} = 2 (\ln \mathcal{L} - \ln \mathcal{L}_0)  
\end{equation}
where $\mathcal{L}$ represents the likelihood of source hypothesis (e.g. presence of an additional source) and $\mathcal{L}_0$ null hypothesis of absence of source. The significance of such detection can be estimated roughly as a square root of the TS value for one degree of freedom  \citep{1996ApJ...461..396M}. The model map of this fit shows some excess at the location of 4FGL J1844-0345. 
We investigated \texttt{gta.find\_sources} with $4\sigma$ threshold to identify additional sources contributing to the excess. We found few sources in the ROI, but one specific source nearest to pulsar than any $4$FGL source with an offset of $0.07^{\circ}$ degree. Subsequently, once more \texttt{gta.fit()} was done. Here, we have deleted the sources with $\mathrm{TS} \leq 0$ or $N_{\text{pred}} \leq 0$ from the model. The updated TS map, now including the new source, caused a significant reduction in the residuals at the pulsar position, which can be seen in Fig.~\ref{fig:tsmaps}(b). Additionally, we presented the On-phase TS map in Fig.~\ref{fig:tsmaps}(a) for comparison.  Finally, we evaluated the extension of the newly identified source using Fermipy's extension framework, the details of which are presented in the Results section.

\subsection{Analysis Results}
For both analyses, we used a binned likelihood approach offered by ScienceTools version 2.2.0 and fermipy version 1.2.0 \citep{2017ICRC...35..824W}. The results from the standard likelihood analysis are discussed first, followed by the results of the pulsar-phased analysis.
We performed localisation, extension, and source-finding analysis across the entire energy range within the ROI to further refine the spatial and spectral modeling. The standard analysis revealed the closest source to our target was 4FGL J$1844.4$-$0345$, with an Offset of $0.153^{\circ}$. This source exhibits such steady, bright emission within the HAWC source extension, with a TS of $1173.37$, and is a GeV pulsar PSR J1844-0346 \citep{Clark_2017}. Despite accounting for the pulsar, residual gamma-ray emission remained, which was not explained by the model. However, no additional point sources with $\sqrt{\mathrm{TS}} \sim 5$ were detected using Fermipy's source-finding algorithm. This outcome is expected, as such residuals would otherwise meet the source detection criterion and be included in the Fermi-LAT 4FGL catalogue. We extracted the SED of 4FGL J1844.4-0345 using \texttt{gta.sed()}. The SED follows a power law with an exponential cut-off, with an explicit cut-off around 20 GeV. We repeated the analysis using data above 20 GeV to explore the potential contribution beyond the pulsar. In this energy range, 4FGL J$1844.4$-$0345$ no longer exhibited any significant emission. Using the \texttt{gta.find\_sources()}, we didn't find any significant residual at the position of the pulsar beyond 20 GeV. We explored the pulsar-phased analysis during the Off-pulse phase of the pulsar to investigate further.

\begin{figure*}[!htb]
    \centering
    \includegraphics[width=0.48\textwidth]{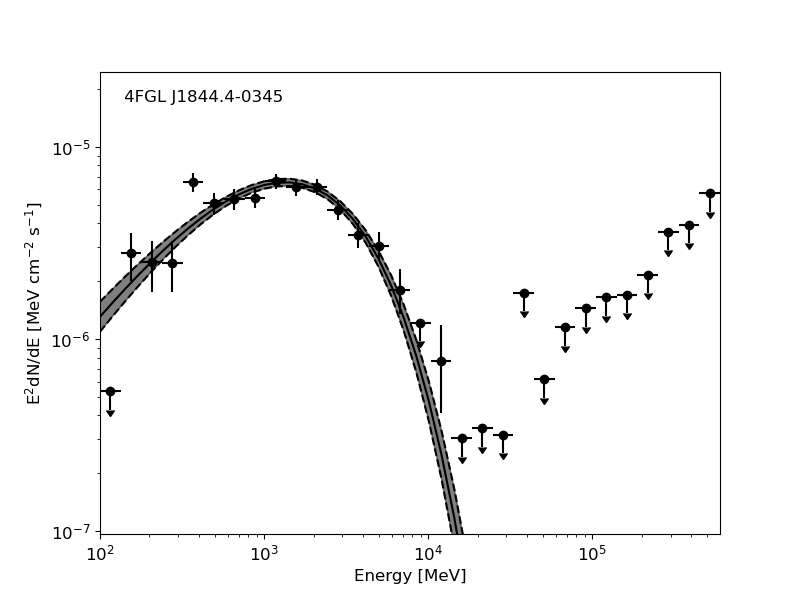}
    \hspace{0.02\textwidth}
    \includegraphics[width=0.48\textwidth]{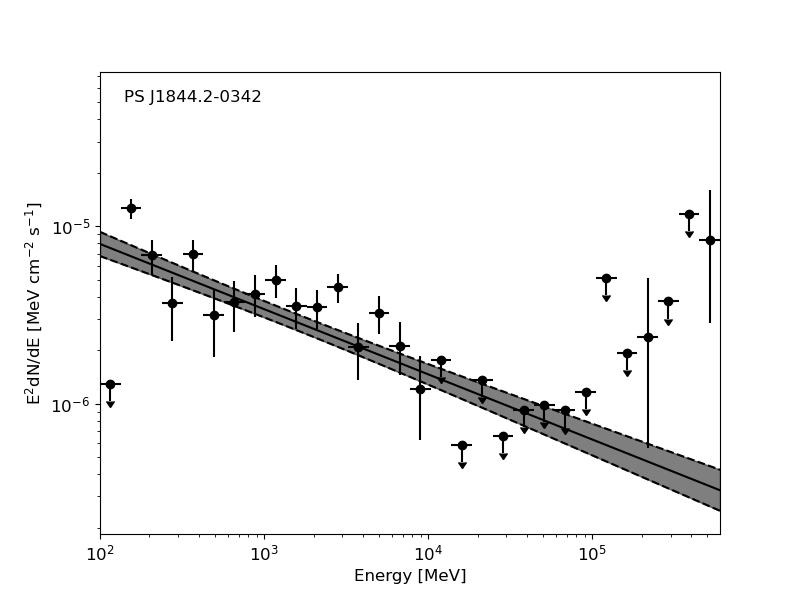}
    \caption{SED's obtained from phased analysis of PSR J1844-0346 (a) On-phase (left panel) (b) Off-phase (right panel)}
    \label{fig:seds}
\end{figure*}


\par The results from the pulsar-phased analysis are summarised here. We found few sources in the ROI with \texttt{gta.find\_sources()}. According to the Fermipy convention, sources found by this method are added to the model and given designations PS JXXXX.X$+$XXXX according to their position in celestial coordinates. In this case, only one specific source named PS J$1844.2$-$0342$ resides within the HAWC extension and is nearer to the pulsar than any $4$FGL source with an Offset of $0.07^{\circ}$ degree. The spectral energy distribution (SED) results are shown in Fig.~\ref{fig:seds}, separated into On-phase (pulsar-dominated) and Off-phase (pulsar wind nebula-dominated) contributions. The On-phase SED is similar to the SED obtained from the standard analysis; both are fitted with power-law with exponential cut-off. The cut-off $\sim 20$ GeV is evident. The error bar represents the $1\sigma$ statistical error, and the confidence band represents the $1\sigma$ error obtained from the covariance matrix. Data points with $5\sigma$ or higher significance from the Off-phase SED are retained for broadband SED modeling, while data points with lower significance, were used to obtain upper limits with $95\%$ confidence level.
The Off-phase phase spectra exhibit a flat power-law profile. In contrast, the On-phase spectrum displays an exponential spectral cut-off, indicative of magnetospheric pulsar emission. We presented the On-phase TS map in Fig.~\ref{fig:tsmaps}(a).The residual in the northern part of Fig.~\ref{fig:tsmaps}(c) after accounting for the new source is outside the extension of our target HAWC source J1844-034 and possibly due to HAWC J1843-032 (\cite{Albert_2023}, see Fig.~6) which we don't explicitly model.

\begin{deluxetable}{ccc}
\tablecaption{Source Properties
\label{tab:source_properties}}
\tablehead{
    \colhead{Spatial template} & \colhead{Best-fit extension (degree)} & \colhead{$\text{TS}_{\text{ext}}$ Significance}
}
\startdata
Radial Gaussian   & 0.172   & 14.32 \\
Radial Disk       & 0.173   & 13.73 \\
\enddata
\end{deluxetable}

\begin{figure*}[!htb]
    \centering
    \includegraphics[width=0.65\textwidth]{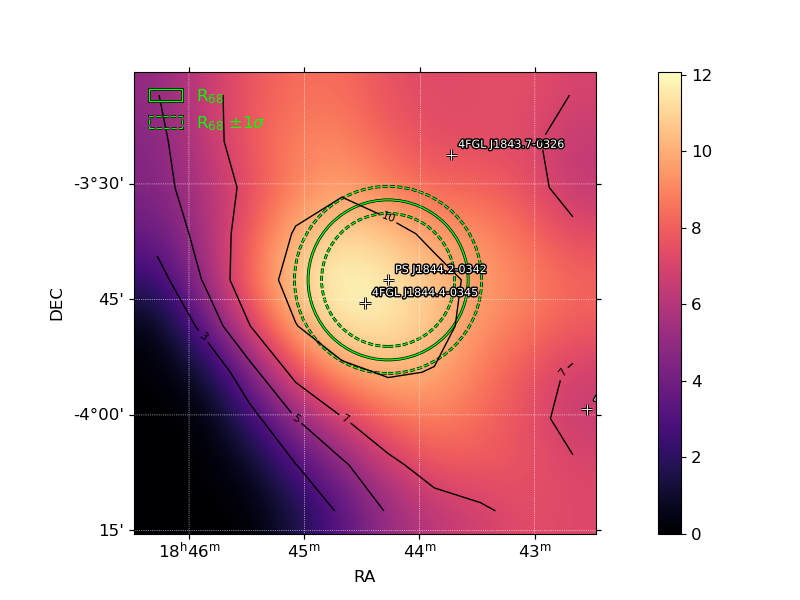}
    \caption{Zoomed-in $\mathrm{TS_{ext}}$  map in full energy range highlighting the extended emission around the pulsar during Off-phase. The $1\sigma$ extension of the newly identified source, PS J$1844.2$-$0342$, is marked in the figure.}
    \label{fig:zoomedtsmap}
\end{figure*}

We used \texttt{gta.extension()} and Fermipy-supported spatial templates to analyse this extended emission. The best-fit parameters, summarised in Table~\ref{tab:source_properties}, favour a radially symmetric Gaussian model with an extension of $0.172^{\circ}$ and a TS extension value of $14.32$, outperforming both radial disk and point-source models. We have also performed the Akaike information criterion (AIC) \citep{Akaike1974ANL,10.1093/mnras/stac3618} to assess further the preferred model between radial disk and radial Gaussian. The AIC is defined as,  $AIC = 2k - 2\ln \mathcal{L}$, where k is the number of free parameters in the model and $\mathcal{L}$ is the maximum likelihood obtained from the best fit. We found that the radial disk model has slightly lower AIC compared to the radial Gaussian; however, the difference in AIC ($\Delta \text{AIC} = 0.44$) is minimal and statistically insignificant, making it inconclusive to favour one model over the other based solely on this criterion. So, we relied on the  $\text{TS}_{\text{ext}}$ criterion to determine the best-fitting model. The extended source, named PS J$1844.2$-$0342$ (RA = $281.068$, Dec = $-3.708$), likely corresponds to a PWN powered by the pulsar. It was detected with TS= $215.9$ and fitted with a power-law spectrum and radially symmetric Gaussian profile. The best-fit power-law spectrum is expressed as $dN/dE = N_0 \left(E/{E_0}\right)^{\gamma}$, where \(N_0\) is the \texttt{prefactor}, \(\gamma\) is the \texttt{spectral index}, and \(E_0\) is the \texttt{energy scale}. The \texttt{prefactor} for the power law spectrum is $4.011 \times 10^{-12} \pm   2.82 \times 10^{-13} \,\text{MeV}^{-1}\text{cm}^{-2}\text{s}^{-1} $ with a \texttt{spectral index} of $\gamma$ = $-2.256 \pm 0.04$ and energy scale $E_0$ of $1$ GeV. The best-fit spectrum is shown in Fig.~\ref{fig:seds}(b). At an assumed distance of $2.4$ kpc, the extension is 20 pc, corresponding to a physical size typical of PWN. A zoomed-in $\mathrm{TS_{ext}}$  map highlighting the extended emission around the pulsar during Off-phase is shown in Fig.~\ref{fig:zoomedtsmap}. We have summarized the key findings of our analysis in Table~\ref{tab:findings}.

\begin{deluxetable*}{|l|p{10cm}|}
\tablecaption{Summary of Analysis Findings\label{tab:findings}}
\tablehead{
\colhead{\textbf{Analysis Type}} & \colhead{\textbf{Key Findings}}
}
\startdata
\text{Standard Analysis} &
\begin{itemize}
    \item The closest identified source was 4FGL J$1844.4$-$0345$, a GeV pulsar with a TS value of 1173.37 and an angular Offset of $0.153^{\circ}$.
    \item The SED of 4FGL J$1844.4$-$0345$ follows a power law with an exponential cut-off, with the cut-off at $\sim$20 GeV.
    \end{itemize} \\
\hline
\text{Pulsar-phased Analysis} &
\begin{itemize}
    \item In the Off-phase of the pulsar, a new source, PS J$1844.2$-$0342$, was identified at an Offset of $0.07^{\circ}$ from the pulsar, located within the HAWC source extension with a high-TS value of 216.
    \item The new source has an extension of $0.172^{\circ}$ with extended hypothesis $\mathrm{TS_{\text{ext}}}$ significance greater than $3\sigma$.
    \item Off-pulse emission suggests contributions from extended sources, potentially a pulsar wind nebula (PWN).
\end{itemize} \\
\enddata
\end{deluxetable*}

\section{PWN SED Modeling with GAMERA}
We assume that the PWN produces the emission observed by HAWC due to the radiative losses by the electrons and positrons being accelerated by its termination shock. The transport equation used to study the time evolution of the relativistic leptons is given by
\begin{equation} 
\frac{\partial{N(E,t)}}{\partial{t}}=Q(E,t)-\frac{\partial{[b(E,t)N(E,t)]}}{\partial{E}}-\frac{N(E,t)}{t_{\text{diff}}}
\end{equation}
 where $N(E,t)$ is the resulting particle spectra at any time t. $t_{\text{diff}}$ is the escape time of high-energy leptons leaving the nebula via Kolmogorov diffusion. $b=b(E,t)$ includes synchrotron and inverse Compton energy losses of the relativistic leptons.  $Q(E,t)$ is the injection spectra. 

\begin{deluxetable}{llcl}
    \tablecaption{Pulsar and Model Parameters\label{tab:params}}
    \tablehead{
        \colhead{Parameter} & 
        \colhead{Description} & 
        \colhead{Value} & 
        \colhead{Units}
    }
    \startdata
    \multicolumn{4}{c}{\textbf{Input Parameters}} \\
    \hline
    $d$         & Pulsar distance$^a$           & 2.4                  & kpc \\
    $\dot{E}$   & Pulsar spin-down power$^b$    & $4.2 \times 10^{36}$ & erg s$^{-1}$ \\
    $\tau_c$    & Pulsar characteristic age$^b$ & 11.6                 & kyr \\
    $P$         & Pulsar period$^b$            & 113                  & ms \\
    $\dot{P}$   & Pulsar period derivative$^b$  & $1.55 \times 10^{-13}$ & s s$^{-1}$ \\
    $n$         & Pulsar braking index$^c$     & 3                    & -- \\
    $E_{\text{min}}$   & Min energy of $e^+ e^-$  & $1.0$               & GeV \\    
    $E_{\text{max}}$   & Max energy of $e^+ e^-$  & $300$               & TeV \\    \hline
    \multicolumn{4}{c}{\textbf{Adjusted Parameters}} \\
    \hline
    $\epsilon$  & $e^+ e^-$ power fraction   & 0.45                  & -- \\
    $B$         & Magnetic field            & 3.4                  & $\mu$G \\
    $P_0$       & Pulsar birth period       & 85                   & ms \\
    $E_c$       & Cut-off energy             & 175                  & TeV \\
    $\alpha$    & Injection spectrum index  & 1.8                  & -- \\
    $t_{\text{age}}$ & Age of pulsar        & 4.7                    & kyr \\
    \enddata
    \tablecomments{Input parameters are taken from observational measurements and theoretical models. Adjusted parameters are fit to the model. $^{a}$ \cite{Wu_2018}, $^b$\cite{Manchester_2005}, $^c$\cite{Tibet}. }
\end{deluxetable}
\par 
We employed the one-zone PWN model, including the energy-dependent modeling of the pulsar energy output, ambient magnetic field and time-dependent injection, following closely the H.E.S.S. collaboration paper \citep{2023}. We modelled the time evolution of particle distribution and radiation SED using GAMERA \citep{2022ascl.soft03007H}. The parameters of the model are of two types: fixed input parameters, whose values come from observations, and adjusted parameters, whose values have been adjusted to fit the SED. The particle injection spectrum followed the power law index with an exponential cut-off. The cut-off energy is treated as a free parameter, and it is denoted as $E_{\text{c}}$. The time evolutions of the pulsar period, pulsar spin-down power, and magnetic field have been used from \cite{Gaensler_2006} and \cite{2007whsn.conf...40V}. The normalisation of the injected particle spectrum is determined by $\epsilon \times \dot{E}(t)$, where $\epsilon$ is $e^+ e^-$ power fraction. The evolution of spin-down luminosity is given by:
\begin{equation}
    \dot{E}(t) = \dot{E}_0 \left( 1 + \frac{t}{\tau_0} \right)^{-2},    
\end{equation}
The time-varying PWN magnetic field is given by :
\begin{equation}
    B(t) = B_0 \left[ 1 + \left( \frac{t}{\tau_0} \right)^{0.5} \right]^{-1}
\end{equation}
where spin-down time scale is given by
\begin{equation}
        \tau_0 = \frac{P_0^2 P ^{- 1}}{2 \dot{P}},
\end{equation}
 The characteristics age of the pulsar is given by $\tau_c = P/{2\dot{P}}$
where $P_0$ and  $\dot{P_0}$ are pulsar birth period and corresponding period derivative respectively. $P$ and $\dot{P}$ means current day pulsar period and their derivative. These two parameter values are known from the pulsar catalogue \citep{Manchester_2005}. We have assumed the birth period of the pulsar as 85 ms, shown in Table~\ref{tab:params}, which gives a value of $\tau_0=6.5$ kyr. 
We adopted the canonical breaking index of pulsar $n=3$. Similar assumptions regarding the birth period and braking index have been made in several studies, such as \cite{10.1093/mnras/stac3618,2023}. The more general form of the above formulas can be found in \cite{Gaensler_2006}. The true age has to be close to 5 kyr, which is the difference between $\tau_c=11.6$ kyr and $\tau_0=6.5$ kyr, following the general relation given in Eqn. (\ref{age_rel}):
\begin{equation}
   \tau_0 = \frac{P_0} {(n-1)\dot{P_0}} = \frac{2\tau_c}{(n-1)}- t_{\text{age}}
\label{age_rel}    
\end{equation}

\begin{figure*}[!htbp] 
\centering
\includegraphics[width=0.65\textwidth]
{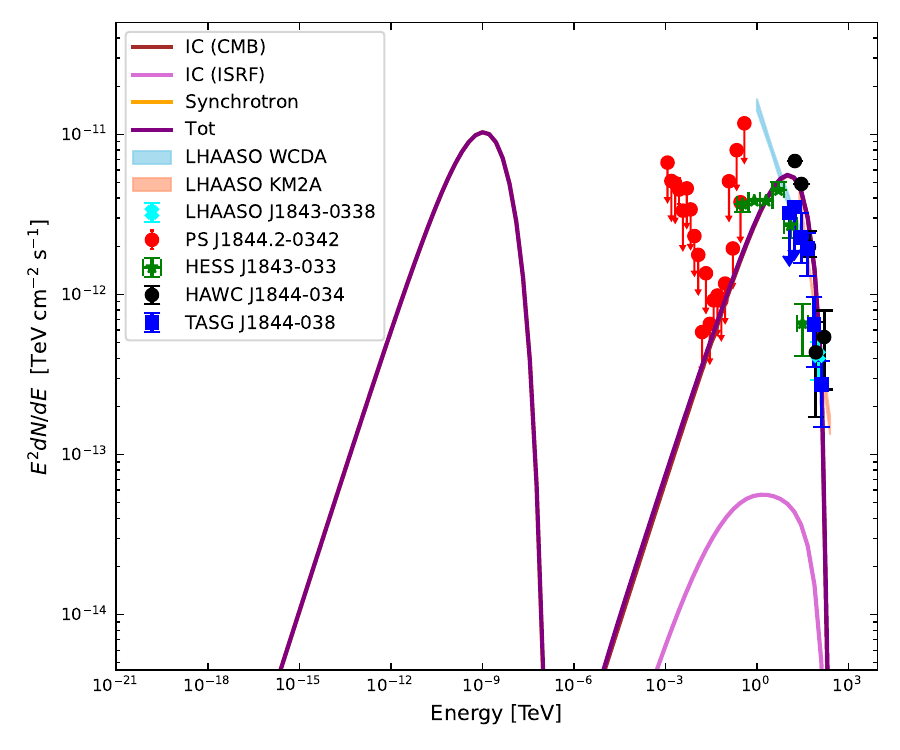}  
    \caption{Gamma-ray data from HAWC \citep{Albert_2023}, LHAASO \citep{UHELHAASO,Cao_2024}, Tibet \citep{Sudoh_2021} and H.E.S.S. \citep{HESScollab2018ana} shown with Fermi-LAT upper limits obtained from our analysis. The SED obtained from our modeling is shown with a solid line.}
    \label{fig:sed}
\end{figure*}

We have computed the radiative cooling losses of the ultra-relativistic electrons due to IC, Synchrotron, and non-radiative loss due to diffusive escape. 
In the GAMERA framework, the complete Klein-Nishina cross-section for IC scattering \citep{RevModPhys.42.237} is utilised to compute the photon flux resulting from the relativistic electrons. For calculating the IC emission from the ISRF target photon field, we have used the stellar and dust contribution in the position of the source as given in \citet{10.1093/mnras/stx1282}. We have included the energy-dependent diffusion loss of cosmic ray electrons using the following form of diffusion coefficient.
\begin{equation} \label{trans}
D = D_0  \Big(\frac{E}{E_0}\Big)^{\delta}   
\end{equation}
where $D_o=1.1 \times 10^{28}$ cm$^{2}$ s$^{-1}$ is the diffusion coefficient normalised at $E_0 = 40$ TeV and $\delta = 0.33$ (Kolomogorov scaling). This delta parameter has a very weak dependence on the observed data. These values are similar to those used for other pulsar halos \citep{Abeysekara_2017}. The values of the parameters used to fit the observational data are presented in Table~\ref{tab:params}, and the SED modeling is shown in Fig.~\ref{fig:sed}. Most of the gamma-ray data points are fitted with the simulated SED obtained from our model, though we do not have any data points at X-ray or lower frequencies to constrain our model. 
It is well discussed in earlier studies that the model parameters cannot be constrained well with the limited number of data points \citep{2023}. We do not have any data points in radio, optical or X-ray frequencies. In the very high energy regime, we have around 13 data points, we tried to fit them by adjusting the values of 6 parameters.
Given some of the model parameters are correlated and not always well constrained from the available data, the derived values of the parameters should not be considered as definitive values. Instead, they represent a plausible combination that provides a reasonable description of the observed data.

\section{Discussion and Conclusion}

The UHE gamma-ray source HAWC J1844-034 is closely associated with several other sources, HAWC J1843-032 and HWC J1846-025. In addition, LHAASO J1843-0338, TASG J1844-038, and HESS J1843-033 are the possible counterparts detected by LHAASO, Tibet AS$_{\gamma}$ and H.E.S.S. respectively. We have analysed the Fermi-LAT data in the region of the HAWC source to find the emission from the extended source. The results of our data analysis are summarized in Table \ref{tab:findings}. The Off-pulsed emission suggests the possibility of the existence of a PWN which could be responsible for the UHE gamma-ray emission. 
In the Fermi-LAT energy band, we have obtained upper limits on the gamma-ray flux which are used to constrain the model parameters while modeling the multi-wavelength SED in Fig.~\ref{fig:sed}. The values of the parameters of the pulsar PSR J1844-0346 are constrained from observations. They are denoted as input parameters in Table \ref{tab:params}. The values of the other parameters, including the pulsar birth period are adjusted to fit the observational data. Our model is consistent with the observational data.

\par
\cite{zhang2024pwn} studied the possibility of TeV gamma-ray emission from the putative PWNe associated with four pulsars. one of these pulsars is PSR J1844-0346. They showed that the TeV emission of the potential PWNe of the pulsars PSR 1838-0537 and PSR J1844-0346 are relatively high compared to those associated with PSR J1208-6238 and PSR J1341-6220. Their results suggested that this TeV emission could be detectable by S-CTA and even by H.E.S.S. for long-duration observations. The X-ray observations from the PWNe would be useful to constrain the models.

\par
\cite{araya2023discoveryextendedgevcounterpart} analysed almost 15 years of data recorded by Fermi-LAT in the region of the TeV gamma-ray source 1LHAASO J1945+2424. The TeV source is more extended than the Fermi-LAT source and the GeV spectrum connects smoothly with the TeV spectrum. A 4$\sigma$ excess was found in their TS map at the location of the source 4FGL J1948.8+2420 as well as at other nearby locations within the extension of the source 1LHAASO J1945+2424. They searched for new sources within their ROI. They found 10 new sources, three of which PS J1945.2+2419, PS J1945.3+2449 and PS J1948.7+2422 are located within 39$\%$ containment region of the WCDA detected source. They replaced the three newly found sources with an extended source modelled by a 2D Gaussian. They used the flux upper limits and data points from this extended source in the Fermi-LAT energy band to model the multi-wavelength SED of 1LHAASO J1945+2424 assuming both SNR and PWN scenarios.

\par
In this work, we have identified a new source PS J$1844.2$-$0342$ near the pulsar PSR J1844-0346, the details are given in Table~\ref{tab:findings}, and further observations are necessary for the confirmation of our result. In, Fig.~\ref{fig:tsmaps}(c) it is shown that SNR G28.6-0.1 is outside the 1$\sigma$ extension region of PS J$1844.2$-$0342$, hence we suggest that PS J$1844.2$-$0342$ is likely to be the PWN of PSR J1844-0346.

\par
Recently, in many papers, the Fermi-LAT data have been analysed to investigate the emission from extended TeV halos \citep{Li_2021, Di_Mauro_2021, Abe_2023, guo2024revisiongevgammarayemission, xiao2024likelydetectiongevgammaray}. Fermi-LAT data analysis is very useful in exploring the GeV counterparts of the TeV sources and subsequently identifying their origin. If both SNR and pulsar are present within the TeV halo, in some cases with the spatial position and morphology of the extended emission in the Fermi-LAT energy band, it would be possible to identify whether the SNR or the pulsar is the more likely counterpart of the TeV halo.

\par
We conclude from our results that the extended GeV source PS J$1844.2$-$0342$ may be the PWN of pulsar PSR J1844-0346, and it is the GeV counterpart of the TeV extended source HAWC J1844-034. We have modelled  HAWC J1844-034 assuming a PWN origin of TeV emission. More observational data at lower frequencies would be helpful to unravel the nature of HAWC J1844-034.

\section*{Acknowledgement}
The authors thank the referee for helpful comments.

\bibliography{reference}{}
\bibliographystyle{aasjournal}

\end{document}